\documentstyle[prl,aps]{revtex}
\newcommand{\be}{\begin{equation}}
\newcommand{\ee}{\end{equation}}
\newcommand{\ba}{\begin{eqnarray}}
\newcommand{\ea}{\end{eqnarray}}

\begin{document}
\draft
\title{Gauge Invariance and Effective Actions in $D=3$ at Finite 
Temperature}
\author{I.J.R~Aitchison$^a$\,,
C.D.~Fosco$^{b}$\thanks{CONICET}\,
\\
{\normalsize\it 
$^a$University of Oxford, Department of Physics, 
Theoretical Physics, 1 Keble Road,\\ Oxford OX1 3NP, 
United Kingdom}\\
{\normalsize\it
$^b$Centro At\'omico Bariloche,
8400 Bariloche, Argentina}}
\date{\today}
\maketitle
\begin{abstract}
For background gauge field configurations reducible to the form
$A_\mu = ({\tilde A}_3, {\vec A}({\vec x}) )$ where ${\tilde A}_3$
is a constant, we provide an elementary derivation of the recently 
obtained result for the exact induced Chern-Simons (CS) effective 
action in $QED_3$ at finite temperature. The method allows us to extend 
the result in several useful ways: to obtain the analogous result for 
the `mixed' CS term in the Dorey-Mavromatos model of parity-conserving
planar superconductivity, thereby justifying their argument for
flux quantization in the model; to the induced CS term for a 
$\tau$-dependent flux; and to the term of second order in 
${\vec A}({\vec x})$ (and all orders in ${\tilde A}_3$) in the effective 
action.   
\end{abstract}

\bigskip

\section{Introduction.}
Recently, there has been significant progress~\cite{dll,dgs,frs1,frs2} 
in resolving a puzzle
concerning the gauge invariance of induced Chern-Simons (CS) terms at
finite temperature, $T$. In the non-Abelian case, for example, general
arguments imply that the coefficient of the CS 
term must be quantized at zero temperature~\cite{djt}, and also
at non-zero temperature~\cite{alvarez,cfrs}, if the action is to
be invariant under topologically non-trivial (`large') gauge
transformations. On the other hand, simple perturbative 
calculations~\cite{NS,N,NS1,DP,DP1,AF,P,Bu,K,I}, 
give the result that the
effect of moving to $T \neq 0$ is simply to multiply the
zero-temperature CS term by a smoothly varying function of T
(typically, ${\rm tanh}(\frac{\beta |M|}{2})$, where 
$\beta = \frac{1}{k T}$ and $M$ is the mass of the fermion(s) in
the theory).
Plainly, the perturbative result contradicts the quantization
requirement~\cite{pis,Kao,Zuk}. A similar difficulty arises in the
Euclidean case for $T\neq 0$ due to the $S^1$ topology of the
compactified Euclidean time.

Apart from its theoretical interest, the resolution of this puzzle
is important in some physical applications. To give one specific
example, consider the Dorey-Mavromatos (DM) model~\cite{dm} 
of two-dimensional superconductivity without parity violation. 
This model employs two $U(1)$ gauge fields, one the electromagnetic 
field $A_\mu$, the
other a `statistical' gauge field $a_\mu$, 
which is also massless.There are $N_f \geq 2$ flavours of 
four-component fermions, the mass term
is parity conserving, and $A_\mu$ and $a_\mu$ have opposite parity.
At zero temperature a `mixed Chern-Simons (MCS) term' is generated by a
fermion loop with one external $A$ and one external $a$ leg, the 
leading contribution to the action, in powers of derivatives, being
\be
\Gamma_{MCS}\;=\; N_f \frac{e g}{2 \pi} \frac{M}{|M|} \int d^3 x 
\epsilon^{\mu\nu\rho} a_\mu \partial_\nu A_\rho \;,
\label{mcs}
\ee
where $e$ and $g$ are the couplings of $A$ and $a$, respectively.
Note that this term is not parity violating, though it is 
`topological'. One major novelty of this model is that it provides 
a mechanism for superconductivity in two space dimensions 
at $T\neq 0$, without the existence of an order parameter which is a non-singlet
under $U(1)_{EM}$ (such a non-zero order parameter would violate 
the Coleman-Mermin-Wagner theorem~\cite{cmw}).
Without such an order parameter, however, it is difficult to
see how some familiar phenomenological features of superconductivity
can arise -in particular flux quantization, which is conventionally
derived from a Ginzburg-Landau Lagrangian, assuming the order
parameter has charge $2 e$ (pair field). In their model, DM
argued as follows. At $T\neq 0$, (\ref{mcs}) becomes
\be
\Gamma_{MCS}(T\neq 0)\;=\; i N_f \frac{M}{|M|} \frac{e g}{2 \pi} 
\int_0^\beta d \tau \int d^2 x \; \epsilon^{\mu\nu\rho} a_\mu 
\partial_\nu A_\rho \;.
\label{mcs1}
\ee
Consider now a configuration in which $a_\mu = (a_3(\tau), {\vec 0})$
and $A_\rho = (0, {\vec A}({\vec x}))$. Its contribution to the
action should be invariant under topologically non-trivial gauge
transformations on $a_3$, of the form
\be
a_3 \; \to \; a_3 \,+\, \partial_\tau \Omega (\tau) 
\label{fq1}
\ee
with $\Omega (\beta)- \Omega (0) = \frac{2 n \pi}{g}$, where $n$
is an integer. Under (\ref{fq1}), the variation of 
$\Gamma_{MCS}(T\neq 0)$ is 
\be
\delta \Gamma_{MCS}(T\neq 0)\;=\; i \,N_f \,\frac{M}{|M|} \,n \, e  
\int d^2 x \epsilon_{i j} \partial_i A_j \;.
\label{fq3}
\ee
Considering then a superconducting annulus enclosing flux $\Phi$, it 
follows that $\Phi$ has to have the value (restoring $\hbar$ and
$c$)
\be
\Phi \;=\; \frac{m\, h \, c}{N_f \, e}
\label{fq2}
\ee
where $m$ is an integer,
if $\delta \Gamma_{MCS}(T\neq 0)$ is to be an integer multiple of
$2 \pi i$, for any $n$ in (\ref{fq3}).
(\ref{fq2}) gives the required result for $N_f=2$, the value indicated in 
the DM model.
             
Thus, the flux quantization does not come from a charge-$2 e$
order parameter in this model, but precisely by requiring 
invariance, under topologically non-trivial gauge transformations,
of a CS-like term.
Unfortunately, however, (\ref{mcs1}) is only correct 
for $T$ infinitesimally close to $0$. Indeed, the standard lowest order
derivative expansion calculation would yield (\ref{mcs1}) multiplied by
${\rm tanh}(\frac{\beta M}{2})$, as stated above, and this 
$T$-dependent factor would destroy the result (\ref{fq2}).

Nevertheless, by analogy with other CS-like terms at $T \neq 0$,
there is reason to think that the quantization result ({\ref{fq2})
should be true after all. Perhaps the blame could be laid on using 
perturbation theory for the field undergoing the large gauge
transformation. In an attempt to get away from perturbation theory,
Cabra et al~\cite{cfrs}, and Bralic et al~\cite{bfs} considered how a 
more general ansatz, in which the ${\rm tanh}$ function is replaced by a 
general function $F(T)$, could be reconciled with gauge invariance,
and concluded that $F$ could only be a discrete-valued function
of the temperature. The arguments of \cite{cfrs} and \cite{bfs}, 
however, depend crucially on the form of the ansatz assumed, 
and it turns out that it does not, in fact, represent the true 
non-perturbative structure.

A crucial advance was made by Dunne et al~\cite{dll}, who considered 
a solvable model in $0+1$ dimensions, in which a CS-like 
term is generated at zero temperature. These authors found
that the {\rm exact} effective action at finite temperature
was indeed gauge invariant, even though its perturbative
expansion produced a result precisely analogous to that found
in $2+1$ dimensions, namely, the gauge-non-invariant form
`${\rm tanh}(\frac{\beta M}{2}) \times \Gamma_{CS}(T=0)$'.
As Dunne et al remark~\cite{dll}, the exact result cannot be written
as the integral of a density, in Euclidean spacetime, suggesting
that the type of ansatz considered in \cite{cfrs} and \cite{bfs} is 
not adequate. 

Subsequently, non-perturbative calculations of the effective
action in the $2+1$ Abelian case~\cite{dgs} and its explicit 
temperature-dependent parity-breaking part~\cite{frs1} have shown 
that the complete effective action is indeed invariant under
both large and small gauge transformations, the result of \cite{frs1},
in particular, showing some remarkable similarity to that
of Dunne et al~\cite{dll}, as we shall discuss further in section III 
below. The calculation of \cite{frs1} has now been
extended to the non-Abelian case \cite{frs2}.

It is important to note, however, that the explicit 
$2+1$-dimensional results have only been obtained for a
particular class of background gauge field configuration.
In the Abelian case, for example, they have the form
\be
A_\mu \;=\; ({\tilde A}_3,{\vec A}({\vec x})) \;,
\ee
where ${\tilde A}_3=\frac{1}{\beta}\int_0^\beta 
d\tau A_3(\tau)$ (these configurations are gauge-equivalent to
those in which $A_\mu \,=\, (A_3(\tau),{\vec A}({\vec x}))$,
as discussed in the following section). 
For this case, the result of \cite{frs1} and \cite{frs2} is the 
following. For a theory with one two-component fermion of mass $M$, 
the zero temperature limit of the induced CS action is
\be
\Gamma_{CS}(T \to 0)\;=\;\frac{i e}{2 \pi} \frac{|M|}{M}
\frac{e \beta {\tilde A}_3}{2} \int d^2 x \epsilon_{j k}
\partial_j A_k \;,
\label{te0}
\ee 
while at finite temperature the result is
\be
\Gamma_{CS}(T\neq 0)\;=\;\frac{i e}{2 \pi} \frac{|M|}{M}
F(\frac{e \beta {\tilde A}_3}{2}) \int d^2 x \epsilon_{j k}
\partial_j A_k \;,
\ee
where
\be
F(x)\;\;=\;\; {\rm arctan}\left[\,
{\rm tanh}(\frac{\beta |M|}{2}) {\rm tan}(x) \,\right] \;.
\ee
Remarkably enough, one sees that $F$ changes by $n \pi$ 
when $A_3$ undergoes a large gauge transformation with
winding number $n$, granted that the branch of the
${\rm arctan}$ is understood to be shifted correspondingly. This is
exactly the same behaviour as in the $T \to 0$ limit (\ref{te0}),
and consequently, quantization arguments are verified for
$T \neq 0$ also, for these configurations. Note  that such
configurations would, in fact, be adequate for analyzing 
the Dorey-Mavromatos flux quantization argument. 

The calculations of \cite{frs1} and \cite{frs2} were formulated 
in such a way as to make  essential use of the result for an 
anomalous Fujikawa Jacobian~\cite{fuji} associated with a global 
chiral rotation on the fermionic variables, involving ${\tilde A}_3$. 
This result is intrinsically non-perturbative, and reveals the
true structure of the odd-parity term, at least for the 
specified configuration. But, while certainly leading to a
satisfactory outcome as regards gauge invariance, the method
of \cite{frs1} and \cite{frs2} seems perhaps rather special, 
and possibly difficult to generalize to other situations, for example, 
the Dorey-Mavromatos problem. In addition, it would be 
interesting to know if the same result could be obtained
by somehow summing up all powers of ${\tilde A}_3$ in a
conventional perturbative (plus derivative) expansion
approach, as actually envisaged by Dunne et al \cite{dll}. After
all, it was in the latter context that the gauge non-invariance
puzzle emerged, and it would be nice to see its resolution
there too.  Such an approach is certainly capable of treating
anomaly-like terms correctly~\cite{af}, provided a suitable regularization
is performed. Non-topological contributions can  be obtained
this way too, of course, and the approach might be capable of
handling more general background field configurations than
those considered in \cite{frs1} and \cite{frs2}. 

The purpose of the present paper is to provide such an
alternative derivation of the results of \cite{frs1} and \cite{frs2}, 
based on straightforward effective action techniques; 
to extend it to the case in which the flux is allowed to
depend on $\tau$;
to derive the 
corresponding result for the DM model, thus rescuing their flux quantization
argument; and to present the result for the non-topological term
which is quadratic in the magnetic field $F_{ij}$, and correct
to all orders in ${\tilde A}_3$, for the case of a single gauge 
field $A$. 

Before proceeding, one important remark needs to be made. The
gauge field configurations which are considered in \cite{frs1} and
\cite{frs2} are effectively $\tau$-independent (or
equivalent to $\tau$-independent ones), and this is crucial in all 
the existing explicit calculations including our own which follow. 
In the non-static {\em time\/}-dependent case, terms arise (due to 
Landau damping) which are non-analytic at the origin of momentum space, 
and which are therefore intrinsically non-local, as has been emphasized 
in \cite{Zuk}. 
Finding the explicit functional dependence on $A_\mu$ which ensures
gauge invariance in the non-static case seems to be difficult.
However, if one assumes
that (after including the gauge field dynamics) only static quantities 
are going to be considered, namely, that
the timelike momentum is always imaginary and discrete, then 
Landau damping cannot occur. This is what happens in the calculation
of static quantities, such as the free energy.

The structure of this paper is as follows: In section II we present
the calculation of the term linear in the magnetic flux, discussing
in particular the cases: a) One two-component fermionic flavour and 
one gauge field, b) Many four-component flavours and two gauge fields 
(mixed CS term), and
c) Extension to the case of a $\tau$-dependent $A_j$. Section III
deals with the next term in the expansion in powers of $A_j$.

\section{Terms linear in the magnetic flux.}
\subsection{Two-component case: parity-breaking term.}
We shall consider here the effective action $\Gamma (A)$ which is induced 
by integrating out a massive two-component fermion field coupled to an 
Abelian gauge field $A_\mu$ in $2+1$ dimensions at finite temperature,
\be
e^{- \Gamma (A)} \;=\; \int {\cal D} {\bar \psi} \,{\cal D} \psi \;
\exp [- S_F (A)] \; .
\ee 
The Euclidean action $S_F (A)$ for the fermion is given by
\be
S_F (A) \;=\; \int_0^\beta d \tau \int d^2 x \; {\bar \psi}
( \not \! \partial + i e \not \!\! A + M ) \psi \; .
\label{sfa}
\ee
In this section, we are using Euclidean Dirac matrices in the
irreducible representation of the Dirac algebra (reducible
representations will be considered in the next section):
\be
\gamma_1 = \sigma_1 \;\;\;\;\gamma_2 = \sigma_2 \;\;\;\;
\gamma_3 = \sigma_3
\ee
where $\sigma_i$ are the usual Pauli matrices
and $\beta = {1}/{T}$ is the inverse temperature.
The label $3$ is used to denote the Euclidean time coordinate $\tau$.
The fermionic fields obey  antiperiodic boundary
conditions in the timelike direction
\be
\psi (\beta , x) \;=\; - \, \psi (0 , x) \;\;\;\;,\;\;\;\;
{\bar \psi} (\beta,x) \;=\; - {\bar \psi} (0, x) \;\;, \forall x \;,
\label{fbc}
\ee
with $x$ denoting the two space coordinates. The gauge field
satisfies periodic boundary conditions
\be
A_\mu (\beta,x) \;=\; A_\mu (0,x) \;\;,\;\; \forall x \;.
\label{gbc}
\ee

We shall first consider configurations satisfying
\be
A_3\;=\;A_3 (\tau) \;\;,\;\; A_j \;=\; A_j(x) ,\;\; j=1,2 \;.
\ee
As stated in ref.~\cite{frs1}, and elaborated in ref.~\cite{frs2}, 
these configurations allow one to study gauge invariance under 
transformations with non-trivial windings around the 
compactified time coordinate:

$$\psi (\tau,x) \; \to \; e^{-i e \Omega (\tau,x)} \psi (\tau,x) \;\;,\;\;
{\bar \psi} (\tau,x) \; \to \; e^{i e \Omega (\tau,x)} {\bar \psi}
(\tau,x)$$
\be
A_\mu (\tau,x) \;\to \; A_\mu (\tau,x) \,+\, \partial_\mu \Omega 
(\tau,x)
\ee
where 
\be
\Omega(\beta,x) \;=\; \Omega(0,x) \,+\, \frac{2 \pi}{e} \, k
\label{omegabc}
\ee
and $k$ is an integer which labels the homotopy class of the gauge
transformation.

As shown in ref.~\cite{frs1}, the $\tau$-dependence of the Dirac operator, 
which comes only from $A_3$, can be removed by a redefinition of the 
integrated fermionic fields with a gauge function
\be
\Omega (\tau) \;=\; - \int_0^\tau d \tau ' A_3 (\tau ') +
\frac{1}{\beta} \int_0^\beta d \tau ' A_3 (\tau ') \;,
\ee 
without affecting the spatial components $A_j$.
Such a transformation renders $A_3$ constant and equal to its mean value,
\be
{\tilde A}_3 \;=\; \frac{1}{\beta} \, \int_0^\beta \, d \tau \, A_3 (\tau)
\;.
\ee
Following \cite{frs1}, the determinant 
is written as an infinite product of the 
corresponding $1+1$ Euclidean Dirac operators
\be
\det ( \not \! \partial + i e \not \!\! A \,+\, M ) \; = \;
\prod_{n=-\infty}^{n=+\infty} \det [\not \! d + M + i \gamma_3 (\omega_n
+ e {\tilde A}_3) ] \;,
\ee
where $\omega_n = (2 n +1) \frac{\pi}{\beta}$ is the usual Matsubara
frequency for fermions and $\not\!\! d$ is the two-dimensional 
Euclidean Dirac operator 
\be
\not \! d \;=\gamma_j (\partial_j + i e A_j) \,=\, 
\not \! \partial + i e \not \!\! A \;,  
\ee 
where we have adopted the convention of using the slash to denote
contraction with the two spatial Dirac matrices, when there is
no risk of confusion.
Then, the effective action $\Gamma (A)$ corresponding to this configuration
will be
\be
\Gamma (A) \;=\; - \sum_{n=-\infty}^{n=+\infty} {\rm Tr} \log 
\left[ \not \! d + i \gamma_3 {\tilde \omega}_n + M \right]
\label{ggen}
\ee
where we have defined ${\tilde \omega}_n\,=\, \omega_n + e {\tilde A}_3$.
This trace cannot, of course, be evaluated explicitly. But if we want
to reproduce the result for the induced parity breaking term of 
\cite{frs1} and \cite{frs2},
it is sufficient to evaluate it up to linear order in $A_j$, without
making any expansion in ${\tilde A}_3$. A naive application of this
approach leads, however, to an ambiguous result, as we will now see. 
Let us call this term $\Gamma^{(1)}(A)$. It is formally given by 
\be
\Gamma^{(1)} (A) \;=\; -i e \sum_{n=-\infty}^{n=+\infty} {\rm Tr} \left[ 
\not \!\! A ( \not \! \partial + i \gamma_3 {\tilde \omega}_n + 
M )^{-1} \right]
\label{gamm1}
\ee 
where the trace is evaluated over functional and Dirac indices.
When written in momentum space, (\ref{gamm1}) becomes
\be
\Gamma^{(1)} (A) \;=\; -i e \sum_{n=-\infty}^{n=+\infty} \,
\int \frac{d^2 p}{(2 \pi)^2} \,{\rm tr}\left[ 
\not \!\! {\tilde A}(0) ( i\not \! p + i \gamma_3 
{\tilde \omega}_n + M )^{-1} 
\right]
\label{naive}
\ee  
where ${\rm tr}$ is the Dirac trace, and
${\tilde A}_j (p)$ is the Fourier transform of $A_j (x)$ with respect 
to the two space variables
\be
{\tilde A}_j (p) \;=\; \int d^2 x \, e^{-i p \cdot x} \, A_j (x) \;.
\label{defft}
\ee

It is immediate to check that, by rationalizing the denominator and
taking the Dirac trace in (\ref{naive}), we obtain $0$, 
which is an unpropitious start to the calculation, and
might seem to contradict the results of reference~\cite{frs1}. The way 
out of this impasse is to realize that, if we are going to deal with 
an $A_j$ such
that the associated magnetic flux is non-zero, then its
zero momentum component is necessarily singular, and the result of 
(\ref{naive}) will be of the ambiguous form $0 \times \infty$.
Indeed, from the definition (\ref{defft}) of the Fourier transform, 
we see that the magnetic flux $\Phi$ can be expressed as the
following limit  
\be
\Phi \;\equiv\; \int d^2 x \, \epsilon_{j k} \partial_j A_k \;=\;
i \lim_{q \to 0} \epsilon_{j k} q_j {\tilde A}_k (q) \;.
\label{flux}
\ee
It follows from (\ref{flux}) that, if we want to have a non-zero
flux~\footnote{Zero-flux configurations give $\Gamma = 0$ without
any ambiguity. This is consistent with the result of \cite{frs1}.}, 
then 
the zero-momentum component 
of ${\tilde A}_j$ necessarily diverges. We somehow need to 
introduce another momentum (`$q$') into the problem, and work with
the finite quantity $\Phi$.

In order to see how to do this let us consider, in fact, the leading 
term in the expansion of $\Gamma^{(1)} (A)$ in powers of 
${\tilde A}_3$, which is just the perturbative action
\be
\Gamma^{(1,1)}(A)\;=\;(i e)^2 \, \sum_{n=-\infty}^{n=+\infty}
{\rm Tr} \left[ \gamma_3 {\tilde A}_3 
(\not \! \partial + i\gamma_3 \omega_n +M)^{-1}
\not \!\! A
(\not \! \partial + i \gamma_3 \omega_n +M)^{-1}\right]\;.
\ee
In this expression it is clear that the trace will involve a
second momentum integration if ${\tilde A}_3$ is allowed to
depend on $x$ as well as on $\tau$:
$$
\Gamma^{(1,1)} ({\tilde A}_3 (x), {\vec A}({\vec x}))\;=\;
(i e)^2 \sum_{n=-\infty}^{n=+\infty} \int d^2 p d^2 q 
{\rm tr} \left[
\langle p+q|\gamma_3 {\tilde A}_3|p\rangle 
(i \not \! p + i \gamma_3 \omega_n + M)^{-1}
\right.
$$
\be
\left. \langle p| \not \! A|p+q\rangle 
(i  (\not \! p+ \not \! q) + i \gamma_3 \omega_n + M)^{-1}\right] \;,
\ee
where ${\rm tr}$ means trace over the Dirac indices only.
The case of $x$-independent ${\tilde A}_3$ may then be obtained
via the limiting process in which the $x$ dependence of
${\tilde A}_3 (x)$ is removed, leading to
$$
\Gamma^{(1,1)}({\tilde A}_3,{\vec A}(x))\;=\;
(i e)^2 {\tilde A}_3 \sum_{n=-\infty}^{+\infty} \,\int
\frac{d^2 p}{(2 \pi)^2} \lim_{q \to 0} {\rm tr} 
\left\{ \gamma_3
(i \not \! p + i \gamma_3 \omega_n + M)^{-1}
\right.
$$
\be
\left.
\not \!\! {\tilde A}(q)
(i  (\not \! p + \not \! q) + i \gamma_3 \omega_n + M)^{-1}
\right\}  \;.
\label{ireg}
\ee
Rationalizing the second denominator and taking the Dirac
trace reveals the presence of a term proportional to $\Phi$,
whose contribution is precisely the odd-parity perturbative 
action, for our special field configuration.
One can proceed in this way, with higher powers of
${\tilde A}_3$, still retaining only one power of ${\vec A}$. 
At the term of order $({\tilde A}_3)^2 {\vec A}$, for example,
one will have three momentum integrals before the $x$-dependence
of ${\tilde A}_3$ is removed, and one might wonder if a further
and more complicated limiting procedure is required.
But it turns out that a well-regulated expression is obtained by
retaining an extra momentum dependence in only the `final'
propagator as in (\ref{ireg}),
$$
\Gamma^{(2,1)}({\tilde A}_3,{\vec A}({\vec x}))\;=\;- (i e)^3
{\tilde A}_3^2 \,\sum_{n=-\infty}^{n=+\infty} \,\int 
\frac{d^2 q}{(2 \pi)^2} \, \lim_{q \to 0} {\rm tr} 
\left\{ \gamma_3 (i \not \! p + i \gamma_3 \omega_n + M)^{-1}
\right.
$$
\be
\left.        
\gamma_3 (i \not \! p + i \gamma_3 \omega_n + M)^{-1}
\not \!\! {\tilde A}(q) 
(i  (\not \! p+ \not \! q) + i \gamma_3 \omega_n + M)^{-1}
\right\} \;,
\label{ireg1}
\ee
and the same is true for higher orders in ${\tilde A}_3$.

But clearly the prospect of evaluating the general term
$\Gamma^{(n,1)}$, and then trying to sum up the answer so
as to obtain the full $\Gamma^{(1)}$, which is non-perturbative
in ${\tilde A}_3$, is unappealing. Fortunately, this is not
necessary. Recall that we resorted to perturbation theory in
${\tilde A}_3$  in order to gain insight into the IR behaviour
-we were not otherwise forced to expand the denominator in
(\ref{naive}) in powers of ${\tilde A}_3$. Indeed, the formulae
(\ref{ireg}) and (\ref{ireg1}) indicate that (\ref{naive}) should
be interpreted in terms of a limit in which the 
argument of the trace is replaced by 
$\not \!\! {\tilde A}(q)(i (\not \! p +\not \! q) + i \gamma_3 
{\tilde \omega}_n + M)^{-1}$, and then the limit $q \to 0$ is taken. 
At this point, however, one realizes that such an expression, 
though IR regular, is UV divergent. 
To regulate the UV divergence, we work instead with
the derivative of $\Gamma (A)$ with respect to ${\tilde A}_3$,
which improves the large momentum dependence of the integrand
by adding an extra propagator, and amounts to subtracting the
value at ${\tilde A}_3 = 0$. These considerations then lead to
the well-behaved expression
\be
\frac{\partial}{\partial {\tilde A}_3} \Gamma (A) \;=\; 
- e^2  \sum_{n=-\infty}^{n=+\infty} \lim_{q \to 0} \, \int d^2 p \,
{\rm tr}
\left\{ \gamma_3 \langle p | (\not \! \partial + i \gamma_3 
{\tilde \omega}_n +M)^{-1} \not \!\!{\tilde A} (
\not \! \partial + i \gamma_3 {\tilde \omega}_n +M)^{-1} 
|p + q \rangle \right\} \;.
\label{ireg2}
\ee

We are now ready to evaluate (\ref{ireg2}). 
After taking the trace, one finds that the only non-vanishing
contribution to the derivative of $\Gamma (A)$ is
\be
\frac{\partial}{\partial {\tilde A}_3} \Gamma (A)\;=\;
2 M e^2 \, \sum_{n=-\infty}^{n=+\infty} \lim_{q \to 0} 
[ \epsilon_{j k} q_j {\tilde A}_k (q)]\int \frac{d^2 p}{(2 \pi)^2}
\frac{1}{(p^2 + {\tilde \omega}_n^2 + M^2)^2}  \;,
\ee 
which, by using (\ref{flux}), may be put as
\be
\frac{\partial}{\partial {\tilde A}_3} \Gamma (A)\;=\;-2 i
M e^2 \, \Phi \, \sum_{n=-\infty}^{n=+\infty} 
\int \frac{d^2 p}{(2 \pi)^2}\frac{1}{(p^2 + {\tilde \omega}_n^2 +
M^2)^2}\;.
\label{mom}
\ee
Performing the (convergent) momentum integral in (\ref{mom}) , we obtain
\be
\frac{\partial}{\partial {\tilde A}_3} \Gamma (A)\;=\;
- \frac{i M e^2}{2 \pi} \; \Phi \, \sum_{n=-\infty}^{n=+\infty}
\frac{1}{{\tilde \omega_n}^2 + M^2} \;.
\ee
When this expression is integrated term by term over ${\tilde A}_3$,
it yields
\be
\Gamma (A)\;=\;
\frac{i e}{2 \pi} \; \Phi \, \sum_{n=-\infty}^{n=+\infty}
{\rm arctan}[\frac{\omega_n + e {\tilde A}_3}{M}]\;.
\label{red1}
\ee
This series is exactly equal to the one that appears in
ref.~\cite{frs1}, and indeed, it shows already the identity of this
result to the one obtained there by the `decoupling' change of
variables. Thus we conclude that, for one fermionic flavour,
and keeping terms linear in $A_j$, the result for the
induced action $\Gamma$ is
\be 
\Gamma (A) \;=\; \frac{i e}{2 \pi} \; \Phi \, {\rm arctan} 
\left[ {\rm tanh} 
\frac{\beta M}{2} {\rm tan} ( \frac{e}{2} \int_0^\beta d \tau A_3 (\tau) ) 
\right] \;.
\label{red2}
\ee 

As in \cite{frs1,frs2}, the branch of the ${\rm arctan}$
in (\ref{red1}) and (\ref{red2}) is defined so as to make $\Gamma$
a continuous function of $e \beta {\tilde A}_3$ as it moves
continuously through the values $2 \pi k$, with $k$ integer.
Also note that, for the non-Abelian configurations considered in
\cite{frs2}, the steps that lead to our final expression (\ref{red2}) 
hold true, with the only addition of a trace over colour indices,
in agreement with \cite{frs2}. 
\subsection{Four-component case: mixed CS term.}
In this section we show what the consequences of the procedure of the 
previous section are for the case of the Lagrangian appearing in the DM model.
In that model, the fermionic action is defined to be
\be
S_f \;=\; \sum_{a=1}^{N_f} \int_0^\beta d \tau \int d^2 x \,
{\bar \psi}_a ( \not \! \partial + i e \not \!\! A + i g \not
\! a \tau_3 + M ) \psi_a 
\ee
where the Dirac matrices are in the reducible $4 \times 4$
representation:
\be
\gamma_\mu \;=\; \left( \begin{array}{cc}
                         \sigma_\mu & 0 \\
                          0 &- \sigma_\mu
                         \end{array} \right)
\;\;\;
\tau_3 \;=\; \left( \begin{array}{cc}
                       I_{2\times 2} & 0 \\
                        0 &- I_{2\times 2}
                       \end{array} \right)  
\;,  
\ee
where $\mu = 1, 2 , 3$, $I_{2 \times 2}$ is the $2 \times 2$
identity matrix and $a$ is the flavour index. It is easy to 
see that the mass term is parity conserving; indeed, rewriting 
the action in terms of two-component fermions, we get
\be
S_f \;=\; \sum_{a=1}^{N_f} \int_0^\beta d \tau \int d^2 x \,\left[
{\bar {\chi_a}_1} ( \not \! \partial + i e \not \!\! A + i g \not
\! a + M ) {\chi_a}_1 \,+\, {\bar {\chi_a}_2} ( \not \! \partial + 
i e \not \!\! A - i g \not \! a - M ) {\chi_a}_2 \right]  
\ee
where ${\chi_a}_1$ and ${\chi_a}_2$ are the upper and lower two-component
spinors corresponding to $\psi_a$.

For the particular configurations $a_3 = a_3(\tau)$, and
$A_j = A_j (x)$ (all the other components $=0$), we can directly 
apply the results of the previous section to each of the two terms 
in the action, just by replacing ${\tilde A}_3$ by ${\tilde a}_3$.
This amounts to using an ${\tilde \omega}_n = \omega_n + e 
{\tilde a}_3$ in the propagators.
As a consequence, for $N_f$ four-component fermions we get
\be
\Gamma (A,a) \;=\; \frac{i e}{\pi} N_f \, \Phi \, 
{\rm arctan} \left[ {\rm tanh} \frac{\beta M}{2}
{\rm tan} ( \frac{g}{2} \int_0^\beta d \tau a_3 (\tau) ) \right] \;,
\ee
which is the result that assures the validity of the flux
quantization argument for any non-zero temperature, as explained in the
Introduction.

\subsection{$\tau$-dependent $A_j$.}
As an example of an extension of the kind of configuration
that can be treated with this method, we shall consider the
case of a gauge field where the constraint of
$\tau$-independence of $A_j$ is relaxed, namely,
\be
A_3 \;=\; A_3 (\tau) \;\;\; A_j \;=\; A_j (\tau,x) \;.
\ee
We will, however, calculate this within the approximation of
keeping terms linear in $A_j$, and everything shall be discussed
for the single flavour two-component case. The necessary changes to make it
appropriate for the many flavours four-component case are analogous to 
the ones performed in section II.B.

As for the previous configurations, we first go to a gauge where
$A_3 (\tau)\to {\tilde A}_3$ becomes a constant. After this, $A_j$ 
will remain  $\tau$-dependent.
In spite of the fact that there is no $\tau$ translation invariance, 
we perform a Fourier transformation with respect to the imaginary time
in (\ref{sfa}), obtaining
\be
S_F (A)\;=\; \frac{1}{\beta} \, \sum_{m=-\infty}^{m=+\infty} 
\sum_{n=-\infty}^{n=+\infty}  \int d^2 x {\bar \psi}_m (x) \left\{
\delta_{m,n} [\not \!\partial + i \gamma_3 {\tilde \omega}_n
+ M]  + i e \not \!\! {\tilde A}^{(m-n)} (x) \right\}  \psi_n (x)
\ee
where
\be
{\tilde A}_j^{(k)} \;=\; \frac{1}{\beta} \int_0^\beta d \tau \,
e^{ - i \frac{2 k \tau}{\beta}} \, A_j (x, \tau) \;.
\ee
The only change that we have to make in the calculation corresponding 
to the static $A_j$ case is that now the matrix whose determinant 
we are evaluating is not diagonal in the space of Matsubara frequencies.
Then
\be
\Gamma (A) \;=\; -{\rm Tr}\log  \left\{
\delta_{m,n} [\not \!\partial + i \gamma_3 {\tilde \omega}_n
+ M]  + i e \not \!\! {\tilde A}^{(m-n)} (x) \right\} 
\ee
where now, of course, the trace also affects the discrete frequencies.
There is an important simplification which occurs because we are
actually dealing with the first order term in $A_j$. When considering
this first order term in the derivative of $\Gamma$ with respect to
${\tilde A}_3$, we obtain
$$
\frac{\partial \Gamma (A)}{\partial {\tilde A}_3} \;=\; 
- e^2  {\rm Tr} \left\{
\gamma_3 (\not \! \partial + i \gamma_3 {\tilde \omega}_m
+M)^{-1} \not \!\!{\tilde A}^{(m-n)} (\not \! \partial + i \gamma_3 
{\tilde \omega}_n +M)^{-1} \right\}
$$
\be
=\;
- e^2  \sum_{n=-\infty}^{n=+\infty} {\rm Tr} \left\{
\gamma_3 (\not \! \partial + i \gamma_3 {\tilde \omega}_n
+M)^{-1} \not \!\! {\tilde A}^{(0)} (\not \! \partial + i \gamma_3 
{\tilde \omega}_n +M)^{-1} \right\}\;.
\ee
Note that only the zero-frequency component of $A_j$ appears in this
expression, which, on the other hand, can now be evaluated analogously
to the static-$A_j$ case, by replacing $A_j$ by its zero-frequency
component. The final result is then:
\be
\Gamma (A) \;=\; \frac{i e}{2 \pi} \; \frac{1}{\beta}\int_0^\beta 
d \tau 
\Phi (\tau)
\, {\rm arctan} \left[ {\rm tanh} 
(\frac{\beta M}{2}) {\rm tan} ( \frac{e}{2} \int_0^\beta 
d \tau' A_3 (\tau') ) \right] \;,
\ee
where $\Phi (\tau) \equiv \int d^2 x \epsilon_{j k}\partial_j A_k (x,\tau)$.

\section{The term quadratic in ${\vec A}$.}
The previous section considered various examples of what is, in fact,
the imaginary part of the full effective action in Euclidean space.
It is this part, linear in the flux, which exhibits the interesting
properties under large gauge transformations. 
The real part of the effective action is not anomalous in that sense, 
and the terms of second and higher order in $A_j$ should be 
straightforwardly gauge invariant. Nevertheless, we think that a 
calculation of the ${\cal O} (A_j^2)$ term has some interest in the
present context, for the following reason. In the soluble 
$(0+1)$ model considered by Dunne et al~\cite{dll} the complete
effective action (for one flavour) is
\be
\Gamma (A) \;=\; \log \left[ 
\cos x \,-\,i {\rm tanh} (\frac{\beta M}{2}) \sin x 
\right]
\label{dunne}
\ee 
where $x=\frac{\beta {\tilde A}}{2}$ and ${\tilde A}
= \frac{1}{\beta} \int_0^\beta d \tau A(\tau)$. The imaginary
part of this is, of course, just the ${\rm arctan}$ function
found in \cite{frs1} and \cite{frs2} and in section II above.
An obvious question to ask is whether the real part of the
action in our $2+1$ case (always for 
our special field configuration) bears any relation to the
real part of (\ref{dunne}). We therefore calculate the
first non-vanishing contribution to the real part, that
of order $A_j^2$, retaining all powers of ${\tilde A}_3$.

We consider one two-component fermion and 
start as before, from the exact expression
for the derivative of the effective action with respect to
${\tilde A}_3$. The term of order two in
$A_j$ (denoted $\frac{\partial \Gamma^{(2)}}{\partial {\tilde A}_3}
(A)$) is
$$
\frac{\partial \Gamma^{(2)}}{\partial {\tilde A}_3} (A)\;=\;
 i e^3 \sum_{n=-\infty}^{+\infty} {\rm Tr} \left\{\gamma_3 
(\not \! \partial + i \gamma_3 {\tilde \omega}_n
+M)^{-1} \not \!\! A (\not \! \partial + i \gamma_3 {\tilde \omega}_n
+M)^{-1} \right.
$$
\be
\left. \not \!\! A (\not \! \partial + i \gamma_3 {\tilde \omega}_n 
+M)^{-1}
\right\}  \;,
\ee    
which needs no IR regularization.      
Evaluating the Dirac trace and the functional trace in momentum space, 
we can write this term as
\be
\frac{\partial \Gamma^{(2)}}{\partial {\tilde A}_3} (A) \;=\;
i e^3 \int \frac{d^2 p}{(2 \pi)^2} \, {\tilde A}_j (p)
\Gamma_{j k} (p) {\tilde A}_k (-p)
\ee
where
\be
\Gamma_{j k} (p)\;=\;2 \sum_{n=-\infty}^{+\infty} i {\tilde \omega}_n
\int \frac{d^2 q}{(2 \pi)^2} \, \left[ \frac{(p^2-q^2-{\tilde \omega}^2_n
- M^2) \delta_{j k} + 4 q_j q_k + 2 
(p_j q_k + q_j p_k)]}{[(p+q)^2 + {\tilde \omega}_n^2 + M^2]^2
(q^2 + {\tilde \omega}_n^2 + M^2)} \right]\;.
\ee
All the momentum integrals appearing in the last expression are
convergent, and moreover, by a lengthy but straightforward calculation we
can recast it into the following explicitly gauge invariant form
\be
\Gamma_{j k} (p)\;=\;2 \sum_{n=-\infty}^{+\infty} i {\tilde \omega}_n
\left[\frac{1}{4\pi} \int_0^1 \frac{d x}{{\cal D}^2_n} (x + x^2 - 2 x^3)
\right] \; (p^2 \delta_{j k} - p_j p_k) \;,
\ee
where
\be
{\cal D}_n \;=\; M^2 + {\tilde \omega}^2_n + x (1-x) p^2 \;. 
\ee
On the other hand, this may also be written as 
\be
\Gamma_{j k} (p)\,=\,\frac{-i}{4\pi e^2 }  
\int_0^1 d x  \frac{x + x^2 - 2 x^3}{\sqrt{M^2 + x(1-x)p^2}}
\frac{\partial^2}{\partial {\tilde A}^2_3} \sum_{n=-\infty}^{+\infty}
{\rm arctan} [\frac{{\tilde \omega}_n}{\sqrt{M^2 + x(1-x)p^2}}]
(p^2 \delta_{j k} - p_j p_k)\,.
\ee
The summation over frequencies can be obtained by borrowing the
result appearing in \cite{frs1,frs2}. Inserting this into 
the expression for the
second order term in the derivative of the effective action,
and integrating over ${\tilde A}_3$, yields
\be
\Gamma^{(2)}({\tilde A}_3,A_j)-\Gamma^{(2)}(0,A_j) \;=\;
i e^3 \int \frac{d^2 p}{(2 \pi)^2} \, {\tilde A}_j (p)
G_{j k} (p) {\tilde A}_k (-p)
\ee
where
$$
G_{j k}(p) \;=\; -\frac{i \beta}{8 \pi e} (p^2 \delta_{j k} -
p_j p_k) \; \int_0^1 d x \,\frac{x + x^2 - 2 x^3}{\sqrt{M^2 + x(1-x)p^2}} 
\times $$
\be
\frac{{\rm tanh}[\frac{\beta}{2}\sqrt{M^2+x(1-x)p^2}]}{
{\rm cos}^2 (\frac{e \beta {\tilde A}_3}{2}) + 
{\rm tanh}^2[\frac{\beta}{2}\sqrt{M^2+x(1-x)p^2}]
{\rm sin}^2 (\frac{e \beta {\tilde A}_3}{2})} \;.
\label{expr}
\ee
We do not write the explicit form of $\Gamma^{(2)}(0,A_j)$
because it is perturbative and insensitive to large gauge transformations. 
Indeed, it can be obtained, for example, by putting $A_3 = 0$ and
$A_j = A_j (x)$ in the result for the induced parity conserving term 
presented in \cite{Kao}. 
The expression (\ref{expr}) would of course become non-local if converted
to the coordinate-space representation; nevertheless it is, in fact,
invariant under large gauge transformations on ${\tilde A}_3$.
A derivative expansion
of (\ref{expr}) gives a series of local terms, the leading one of which 
is

\be
\Gamma^{(2)}({\tilde A}_3,A_j)-\Gamma^{(2)}(0,A_j)
\;\simeq\; \frac{e^2 \beta}{48 \pi M}\;
\frac{{\rm tanh}(\frac{\beta M}{2})}{
{\rm cos}^2 (\frac{e \beta {\tilde A}_3}{2}) + 
{\rm tanh}^2 (\frac{\beta M}{2})
{\rm sin}^2 (\frac{e \beta {\tilde A}_3}{2})}
\int d^2 x F_{j k} F_{j k} \;. 
\label{qq}
\ee

It is indeed amusing, and perhaps of some significance, that
the denominator function appearing in (\ref{qq}) is just the modulus 
squared of the complex function whose logarithm is the result 
(\ref{dunne}) of Dunne et al~\cite{dll}, just as the imaginary part 
of our effective action involves the phase of that function. We have, 
however, not been able to explore this possible connection any further
as yet.

\newpage
\underline{Acknowledgements}: 

This work was performed while I.J.R.A. was visiting Instituto
Balseiro and Centro At\'omico Bariloche, Argentina. I.J.R.A.
is very grateful to the Argentine National Academy of Sciences
for the invitation to visit Argentina, and to the British Council 
for a grant to cover the travel cost. He is also grateful to the
National Academy of Sciences, to the Physics Department of 
La Plata University, and to Instituto Balseiro for their
financial support during the visit.
He acknowledges with particular thanks the warm hospitality
of Dr. H.~Fanchiotti, Dr.. C.~Garcia Canal, and Dr. 
F.~A.~Schaposnik at La Plata University, and of his co-author and the
other members of the Centro At\'omico at Bariloche.

\end{document}